\documentclass[11pt]{article}
\newcommand{\be}{\begin{eqnarray}}\newcommand{\beq}{\begin{equation}}
\newcommand{\ee}{\end{eqnarray}}\newcommand{\eeq}{\end{equation}}

\newcommand{\De}{\Delta}
\input epsf 
\usepackage{graphicx} 
\usepackage{amssymb}
\usepackage{wasysym}
\title{
Dependence of homogeneous crystal nucleation in water droplets on their radii and its implication for 
modeling the formation of ice particles in cirrus clouds 
}
\author{Yuri S. Djikaev\thanks{Corresponding author. E-mail: idjikaev@buffalo.edu}\hspace{0.2cm} 
and \hspace{0.1cm} Eli Ruckenstein$^{}$\thanks{
E-mail: feaeliru@buffalo.edu }\hspace{0.2cm} \\ 
\\ Department of Chemical and Biological  Engineering, SUNY at Buffalo, \\ 
Buffalo, New York  14260 }
\date{ \hfill }
\renewcommand{\baselinestretch}{2}
\begin{document}
\renewcommand{\baselinestretch}{1}
\maketitle
\renewcommand{\baselinestretch}{1}
{\bf Abstract.} 
{\small

We propose an approximation for the total ice nucleation rate $J=J(T,R)$ in supercooled water  droplets as a
function of  both droplet radius $R$ and temperature $T$, taking account of both  volume-based and
surface-stimulated nucleation modes. Its crucial idea is that, even in the 
surface-stimulated mode crystal nuclei initially emerge (as sub-critical clusters) homogeneously in the 
sub-surface  layer, not ``pseudo-heterogeneously"  at the surface. This mode is negligible in large droplets, 
but becomes increasingly  important with decreasing droplet size and is dominant in small droplets. The
crossover droplet radius for the transition of homogeneous ice nucleation from the volume-based mode to the
surface-stimulated mode nonmonotonically depends on $T$ from $233\;$K to $239.5\;$K, ranging there from 
$\approx2\;\mu$m to $\approx100\;\mu$m.  Using experimental data on ice nucleation rates in small droplets, 
we determine that the ice-air interfacial tension of basal
facets of ice crystals monotonically increases from $\approx89$ dyn/cm at $T=234.8\mbox{ K}$ to
$\approx91$  dyn/cm at  $T=236.2\mbox{ K}$.

} 
\begin{figure}[h]
\begin{center}\vspace{0.3cm}
\includegraphics[width=6.5cm]{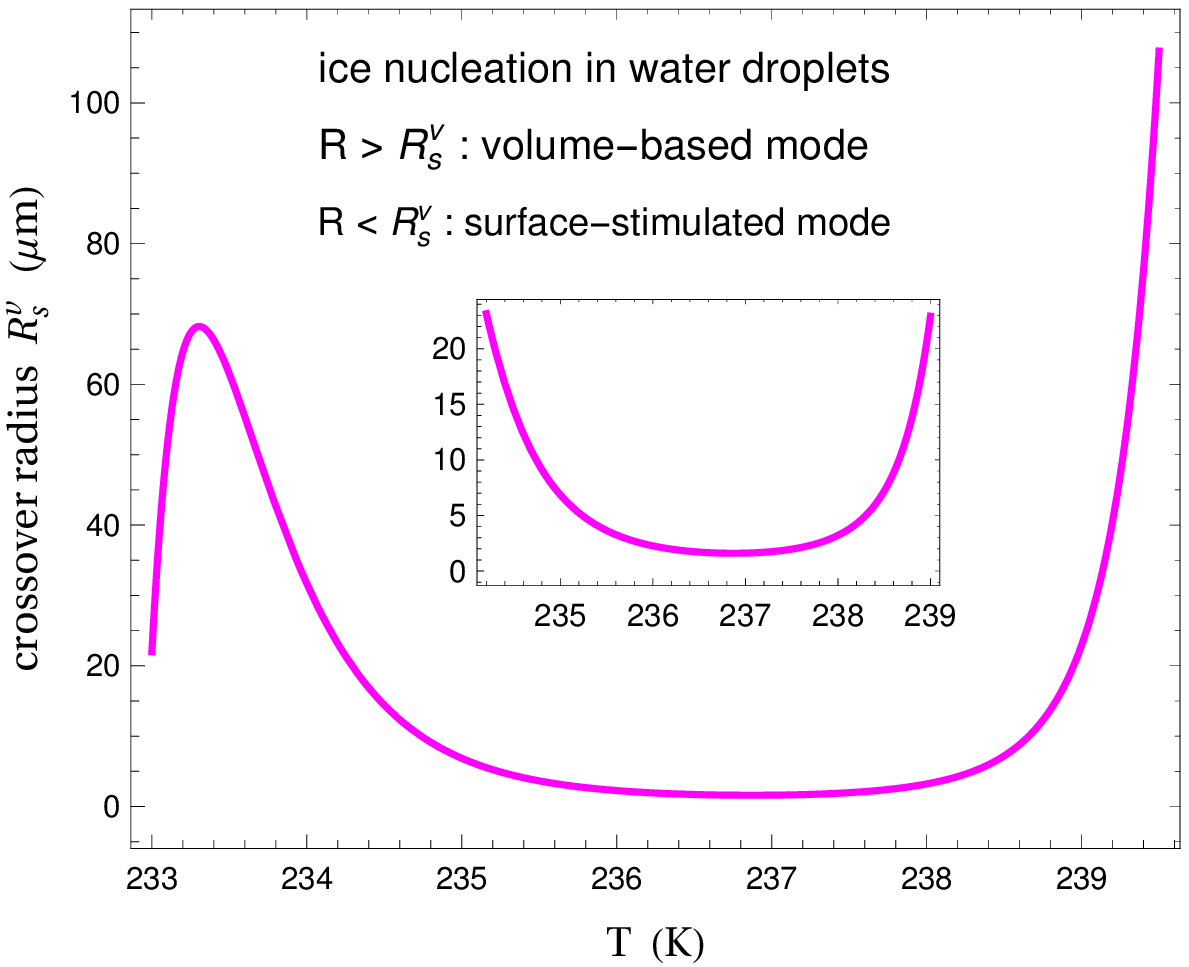}\\ [3.cm]
\vspace{-3.2cm}
\end{center}
\end{figure} 
\renewcommand{\baselinestretch}{1} 
\newpage 
\section{Introduction}
\renewcommand{\baselinestretch}{2}

Water is the most widespread and important chemical compound on Earth and hence, not surprisingly,  most
intensively studied.$^{1,2}$ In the atmosphere, water is present in both vapor and condensed phases  
(liquid or solid aqueous particles).$^{3,4}$  Atmospheric aerosol and cloud particles
influence Earth's global climate in various ways, e.g., by affecting the 
solar and terrestrial radiation
balance, providing heterogeneous/catalytic surfaces for chemical reactions of some atmospheric species,
removing some chemical compounds from atmosphere via precipitations, etc... 
These effects depends on 
the radiative and other physical and chemical properties of the aerosol and cloud particles.$^{4}$ 
Therefore, it is necessary for both global and regional climate models to adequately predict the
evolution of physico-chemical properties of atmospheric aerosols depending on  
atmospheric conditions. 

One of the most important processes drastically affecting the radiative and catalytic properties of water
particles in the atmosphere is the freezing of supercooled liquid droplets into ice.$^{5}$  This 
phenomenon constitutes the core of the formation of wispy cirrus clouds (typically found  at heights above 6
km).$^{6}$  Since the radiative properties of ice and its ability to act as catalytic surface for
heterogeneous chemical reactions differ drastically from those of liquid water, the formation of cirrus clouds
is accompanied by  significant changes in Earth's radiative budget and possible heterogeneous chemical
reactions involving  atmospheric gaseous species.$^{7,8}$ Thus it is important to be able to predict under what
set of environmental conditions supercooled water droplets freeze into ice. 

If no foreign (heterogeneous) centers are present in liquid water,  it can be {\em supercooled} to
temperatures  below the melting temperature of ice and remain liquid for some time.  For the crystallization
process to start, an ice-like cluster of critical size (crystal nucleus)  must first appear in water.
Without heterogeneous centers,   the formation of a crystal nucleus is a very rare event.  Such
nuclei do form due to density and structure fluctuations in liquid; this process is called nucleation 
(heterogeneous nucleation if heterogeneous centers  are present in liquid, otherwise homogeneous nucleation).

Although many phase transformations in aerosols and droplets occur via heterogeneous nucleation,$^{4}$ in a
number of important cases atmospheric particles appear to freeze homogeneously.$^{9-13}$ The  homogeneous
freezing of supercooled aqueous droplets at temperatures below about -30$^{o}$C can occur anywhere in the
atmosphere from the surface layer (resulting in ice fogs$^{14}$) to the upper troposphere (in cirrus clouds);
such low temperatures are required for the freezing of water aerosols/droplets when there are not enough
preexisting centers in the air capable of nucleating ice.$^{9-11}$ 

The crystal nucleation rate is one of the most important characteristics of crystallization. It constitutes an
integral element of any regional or global climate model. In what follows, we will consider homogeneous
crystallization of pure water droplets and propose an analytical expression for the  crystal nucleation rate
in water droplets as an explicit function of the droplet radius and temperature.    Although for bulk water
the temperature dependence of the homogeneous crystal nucleation rate  has been studied relatively well (see,
e.g., ref.15 and references therein), this dependence for water droplets will be shown to significantly
differ from the bulk one. Our approach allows one to also evaluate  the ice-air interfacial tension
using experimental data on the rate of  homogeneous ice nucleation in water droplets.  

\section{Volume-based and surface-stimulated modes of homogeneous crystal nucleation in droplets}

Until 2002, homogeneous crystallization in
liquids had been  assumed to initiate within the volume of the supercooled liquid,$^{4,16,17}$ even when the
latter was in the dispersed state (aerosols or droplets); the role of the droplet surface in its
crystallization had been ignored.  However, after re-examining some laboratory data on the homogeneous
freezing of  aqueous nitric acid droplets, Tabazadeh {\em et al.}$^{18}$  suggested crystal nucleation to
occur ``pseudo-heterogeneously"  at the air-droplet interface. Similar conclusions were drawn after
re-analyzing experimental  data on the freezing of pure water droplets.$^{19}$   Moreover, using  the
classical nucleation theory (CNT), Djikaev {\em et al.}$^{20,21}$ developed a thermodynamic theory of
surface-stimulated crystal nucleation. It  prescribes the condition  under which the  surface of a
droplet can stimulate crystal nucleation therein so  that  the formation of a crystal nucleus with one of its
facets at the droplet surface (``surface-stimulated" mode) is thermodynamically favored over its formation
with all the facets {\em within} the liquid phase (``volume-based" mode).   

\subsection{Two contributions to the kinetics of crystal nucleation in droplets}

For both  unary and multicomponent droplets, this condition coincides with the  condition for the partial
wettability of at least one of the crystal nucleus facets by the liquid.$^{4,16,20-23}$ 
Neglecting the contributions from line tensions,  the criterion for whether crystal nucleation in a 
supercooled droplet is or is not thermodynamically stimulated by the surface has the form$^{20,21}$ 
\beq \sigma^{sv}_{\lambda}-\sigma^{lv}<\sigma^{ls}_{\lambda},\eeq
where $\sigma^{lv}$ is the liquid-vapor surface tension, whereas 
$\sigma^{ls}_{\lambda}$ and $\sigma^{sv}_{\lambda}$
are the interfacial tensions of crystal facet $\lambda$ in the liquid and in the vapor, respectively. 
This effect was experimentally observed for  several systems,$^{24-26}$ including 
water-ice$^{27}$ at temperatures at or  below 0$^{o}$C.

Taking this into account, it was suggested$^{18,19}$ that the total crystal nucleation rate $J$ in
experiments on droplet freezing should contain two contributions, one due to the volume-based mode of crystal
nucleation, where a crystal nucleus forms with all the facets {\em within} the liquid phase,   and the other
to the surface-stimulated mode, where one of the facets of a crystal nucleus  constitutes a part of the
droplet surface.   The latter mode can become important when the crystallizing liquid   is in a dispersed
state (which is the case with the freezing of atmospheric aerosols and droplets and many experiments); 
since smaller droplets have a higher surface-to-volume ratio than larger ones, surface-stimulated crystal
nuclei will appear more often in the former  than in the latter (or in bulk). 

However, it was argued$^{28}$ that the formation of a crystal nucleus with one of its facets at the   droplet
surface  {\em cannot start} preferentially {\em at} the surface, because  the latter does not have any sites
which would make the ordering of the surrounding {\em surface}  molecules  thermodynamically more favorable
than the ordering of interior molecules. On the contrary, the crystal surface layer remains disordered far
below the melting   temperature due to weaker constraints on surface-located molecules which have a reduced
number of neighbors and hence a higher vibrational  amplitude compared to bulk ones. Consequently, a thin 
disordered layer forms on the crystal surface at  temperatures far below the melting one. This phenomenon 
(premelting)  has been well established both  experimentally and via molecular dynamics
simulations  (see ref.29 and references therein). Moreover, it was experimentally observed$^{30,31}$   that
the premelting of the (0001) face of hexagonal ice occurs at about $200$ K, far below the lowest temperature
reported for the homogeneous freezing of atmospheric droplets.  

Although a crystal nucleus with one facet at a droplet-vapor interface {\em cannot begin} its formation (as a
subcritical crystal) at the droplet surface,$^{}$ the surface of the droplet {\em can} still  stimulate
crystal  nucleation therein (under condition (1)), but  a crystal cluster  has to emerge  {\em homogeneously}
in a spherical layer adjacent to the droplet surface. When this crystal becomes large
enough (due to density and structure fluctuations), one of its facets hits the droplet surface
and  becomes a nucleus owing to a drastic change in its thermodynamic state.$^{28}$   Any  crystalline
cluster, starting its evolution with its center  in the ``surface-stimulated nucleation" layer, has a
potential to become a nucleus (by means of  fluctuations)  once one of its facets, satisfying eq.(1), meets
the droplet surface. Originally proposed in ref.27, this idea was verified in later
experiments$^{32}$ and molecular dynamics simulations.$^{33}$ Recently we have furthered this idea and
applied it to the freezing of aqueous nitric acid droplets.$^{34,35}$

In this model, the total rate of crystal nucleation as the sum of the contributions from both 
volume-based and surface-stimulated modes, can be written as$^{28,34,35}$ 
\beq J=\Big[1+\big(1-(1-(h_{\lambda}/R)u)^3\big)
\Big(\mbox{e}^{-\frac1{2}(u-1)W_{*}/k_BT}-1\Big)\Big] J_v
\eeq
where the parameter $u=(\sigma^{sv}_{\lambda}-\sigma^{lv})/\sigma^{ls}_{\lambda}$ and  
$h_{\lambda}$ is the height of a pyramid whereof the apex is at the center of the volume-based crystal
nucleus and the basis is the facet $\lambda$, satisfying eq.(1); 
$J_v$ is the rate of volume-based crystal nucleation. In the framework of CNT, 
\beq J_v=\frac{k_BT}{h}\rho_l\mbox{e}^{-\Delta G_{d}/k_BT}\mbox{e}^{-W_{*}/k_BT}, \eeq 
where the density of molecules in the droplet is assumed to be uniform up to the  dividing  surface, and
$W_{*}$ is the work of formation of a  volume-based crystal nucleus.

\section{The dependence of the total rate of homogeneous ice nucleation in droplets on their radii and
temperature}
urrent atmospheric models assume  that homogeneous crystallization of an aerosol or a droplet always 
initiates  therewithin$^{4,16,17,22}$ at the nucleation rate $J_v$, thus neglecting the  ability of the
droplet surface to stimulate crystal nucleation. In order to take into account this effect and  thus increase
the accuracy of atmospheric models (in their predictions of  aerosol/droplet freezing events), it is necessary
to treat the crystal nucleation in a droplet as potentially  occurring due to both volume-based and
surface-stimulated modes, with the combined rate $J$ given by eq.(2). 

The combined crystal nucleation rate $J$ depends on both droplet radius $R$ and temperature $T$.  Implementing
it into an atmospheric model requires $J$ to be a known function of both $T$ and $R$:  $J=J(T,R)$. While the
dependence of $J$ on $R$ is simple, it depends on $T$ not only explicitly, but also implicitly via the 
parameters $u$ and quantities $h_{\lambda}$, $W_*$, and $J_v$. 

For the $T$ dependence of the  volume-based  crystal nucleation rate, $J_v$,  Koop and Murray$^{15}$  
proposed a simple and very accurate approximation. They constrained the key terms in CNT for crystal
nucleation (diffusion activation energy and  ice-liquid interfacial energy)  with physically consistent
approximations of all involved quantities. The only adjustable parameter in their  model is the water-ice 
interfacial tension at one temperature which is determined by fitting CNT predictions  to laboratory data on
the rate of homogeneous ice nucleation between 233.6 K and 238.5 K.  They obtained: 
\beq J_v(T)=10^{x(T)} \;\;\;[\mbox{cm}^{-1}\mbox{s}^{-1}],\eeq 
where the function $x(T)$ is the sixth-order polynomial in $T-T_m$ with known coefficients (see eq.(A9a)
and Table VII in ref.15) and $T_m=273.15$ K is the melting temperature of ice.

The $T$-dependence of $W_*$ can be determined via a classical expression$^{4,22,36}$ 
\beq W_*=W_*(T)=\frac{(16\pi/3)[\sigma^{ls}(T)]^3 [m_{\mbox{w}}/\rho_{\mbox{i}}(T)]^2}
{\left[ \Delta q(T)\ln(T/T_m)\right]^{2}.}, \eeq 
where $m_{\mbox{w}}/\rho_{\mbox{i}}(T)=v_{\mbox{i}}(T)$ is the volume per molecule of water in ice, 
$m_{\mbox{w}}$ is the mass of a water molecule, and $\rho_{\mbox{i}}(T)$ is the mass 
density of ice (given, e.g., in Table I of
ref.15);  for the water-ice interfacial tension $\sigma^{ls}(T)$ and enthalpy of
crystallization  of water $\delta q(T)<0$ per molecule as functions of $T$ one can 
use the approximations also from ref.15 (see eqs.(A1)-(A3) and Table III therein). 

Note that eq.(5) assumes that the crystal nucleus approximately has a spherical shape with the
effective$^{16,20}$ 
water-ice interfacial tension $\sigma^{ls}$ (or, equivalently, that all 
the facets of an ice nucleus have the same interfacial water-ice tension, $\sigma^{ls}_{}$). 
Accordingly, this assumption is implied in eq.(4).$^{}$

Using eq.(18) from ref.21, one can show that, independent of the shape of the crystal nucleus, its
volume $V_*$ as a function of $T$ can be found as 
\beq 
V_*=V_*(T)=2W_*(T)v_{\mbox{i}}(T)\left[ \Delta q(T)\ln(T/T_m)\right]^{-1}.  
\eeq
Adopting a more realistic assumption (compared to the assumption of sphericity) that  the ice nucleus
has a shape of a right hexagonal prism with an aspect ratio $\kappa$ and assuming that the basal facets of
the prism are those that satisfy the criterion of partial wettability, eq.(1), the quantity $h_{\lambda}$
(defined above)  will be equal to the half-height of the prism. Thus, one can obtain for its $T$-dependence 
\beq h_{\lambda }=h_{\lambda }(T)=[\kappa^2V_*(T)]^{1/3}/\sqrt{3}.\eeq 

For large enough  droplets of radii $R\rightarrow\infty$, the RHS of eq.(2) reduces$^{28,34,35}$ to  $J_v$. 
Thus, if crystal nucleation rates $J^{\mbox{\tiny exp}}_{\mbox{\tiny Large}}$ 
are measured in experiments on large droplets,  $J^{\mbox{\tiny exp}}_{\mbox{\tiny Large}}=J_v$. 
If, under identical
experimental conditions, crystal nucleation rates $J^{\mbox{\tiny exp}}_{\mbox{\tiny small}}$ 
are measured in experiments on small 
droplets, 
then the rates 
$J^{\mbox{\tiny exp}}_{\mbox{\tiny small}}$ and $J^{\mbox{\tiny exp}}_{\mbox{\tiny Large}}$ 
will be related as 
\beq \frac{J^{\mbox{\tiny exp}}_{\mbox{\tiny small}}}{J^{\mbox{\tiny exp}}_{\mbox{\tiny
Large}}}= \Big[1+\big(1-(1-(h_{\lambda}(T)/R)u)^3\big)
\Big(\mbox{e}^{-\frac1{2}(u-1)W_{*}(T)/k_BT}-1\Big)\Big].\eeq 

Since the functions $W_{*}(T)$ and $h_{\lambda}(T)$ can be considered to be known (as outlined above), 
equation (8) can be solved with respect to $u$ at any temperature $T$, so that one can build an  approximation
for the function $u=u(T)$. Since (at any given $T$ and $R$) the RHS of eq.(8), i.e., the ratio $f(u)\equiv
J/J_v$,  is a  nonmonotonic function$^{34,35}$ of $u$, equation (8) can formally have two roots: one, $u'$, 
with $df(u)/du|_{u'} > 0$, and the other, $u_0>u'$,  with $df(u)/du|_{u_0} < 0$.  However, the root $u'$ is
not physically meaningful. Indeed, at constant  $\sigma_{lv}$ and $\sigma_{\lambda }^{ls}$, one can expect
$J^{\mbox{\tiny exp}}_{\mbox{\tiny small}}$ to  decrease with increasing $\sigma_{\lambda }^{sv}$ hence,
$df(u)/du$ must be negative. 

As clear from eq.(2), knowing the functions  $W_{*}(T)$, $h_{\lambda}(T)$, and $u(T)$, and using eq.(4), 
one can parameterize 
the total rate of crystal nucleation $J$ (containing the contributions from both 
volume-based and surface-stimulated modes) as a function of both droplet radius $R$ and temperature
$T$:
 \beq J=J(T,R)=\Big[1+\big(1-(1-(h_{\lambda}(T)/R)u(T))^3\big)
\Big(\mbox{e}^{-\frac1{2}(u(T)-1)W_{*}/k_BT}-1\Big)\Big] 10^{x(T)}.\eeq 
This expression can be expected to more adequately describe the freezing of water droplets in cirrus clouds
and thus improve the accuracy of both regional and global climate models. 

\subsection{Temperature dependence of the ice-air interfacial tension}

The $T$-dependence of the parameter $u$ (according to its definition) is determined by the
$T$-dependence of 
$\sigma^{ls}_{\lambda}$, 
$\sigma^{sv}_{\lambda}$, and 
$\sigma^{lv}_{}$. 
The function $\sigma^{ls}_{\lambda}(T)$ can be approximated 
as in ref.15, whereas for $\sigma^{lv}_{}(T)$ one can use an analytical 
fit of experimental data reported in ref.37.$^{}$ Thus, 
one can determine the ice-air interfacial tension $\sigma_{\lambda }^{sv}$ of the
facet $\lambda$ as a function of temperature,
\beq \sigma_{\lambda }^{sv}(T)=\sigma^{lv}(T)+u_{\mbox{\tiny }}(T)\sigma_{\lambda }^{ls}(T).\eeq

This method for determining the function $\sigma_{\lambda}^{sv}(T)$ requires the 
experimental data 
for ice nucleation rates in large and small droplets,  $J^{\mbox{\tiny exp}}_{\mbox{\tiny Large}}$
and $J^{\mbox{\tiny exp}}_{\mbox{\tiny small}}$, to be for identical experimental conditions, droplets
differing only in their size. Otherwise, the method  provides only rough estimates for 
$\sigma_{\lambda }^{sv}$.

\section{Numerical Calculations}

To accurately apply the proposed method for determining the dependence of the total ice nucleation 
rate $J$ in droplets on their radii $R$ and temperature $T$, 
experimental data used in eq.(8) for ice nucleation rates in large and small droplets,  $J^{\mbox{\tiny
exp}}_{\mbox{\tiny Large}}$ and $J^{\mbox{\tiny exp}}_{\mbox{\tiny small}}$, must be for identical
experimental conditions, droplets differing only in their size.  Since such ``ideal" data are not available,
for ice nucleation in {\em large} droplets  we used eq.(4) and assumed  $J^{\mbox{\tiny exp}}_{\mbox{\tiny
Large}}\approx J_v$.$^{}$   For ice nucleation in {\em small} droplets, we  used experimental data$^{38}$ for 
droplets of radii $R\approx 0.5,\; 1.7$, and $2.9$ $\mu$m at different $T$'s. 

Alternatively, if there are two sets of experiments on the freezing of water droplets, 
one set for  droplets of radius $R_1$ and the other for droplets of radius $R_2$, with  $J^{\mbox{\tiny
exp}}_{\mbox{\tiny R$_1$}}$ and $J^{\mbox{\tiny exp}}_{\mbox{\tiny R$_2$}}$ being the respective ice 
nucleation 
rates,  an approximation for the function $u(T)$ can be obtained by numerically solving the equation 
$$\frac{J^{\mbox{\tiny exp}}_{\mbox{\tiny R$_1$}}}{J^{\mbox{\tiny exp}}_{\mbox{\tiny
R$_2$}}}=\frac{f(u)|_{T,R_1}}{f(u)|_{T,R_2}} $$
(t experimental temperatures $T$, instead of solving eq.(8).  Again, in the ideal case, 
experiments for both $R_1$ and $R_2$ should be at identical experimental conditions. 

Note that the aspect ratio $\kappa$ entering eq.(10) via eqs.(7) and (8), is not a variable quantity.  The
thermodynamics and kinetics of crystal nucleation constrain the size and shape of the ice  nucleus with a
unique, ``native" aspect ratio $\kappa_*$,  determined by the conditions that the free energy of formation  of
a crystal cluster have an extremum (saddle point) for the nucleus and by Wulff's relations.$^{4,16,20-22,28}$
Although  the morphology of tiny ice crystal nuclei in liquid water has not been investigated thoroughly so
far, some interesting studies have been reported (see, e.g., ref.39 and ref.4, pp.145-161 and references
therein).      Equilibrium (hexagonal) ice crystals are most likely to have the shape of a  regular hexagonal
prism. 
Using the estimates for the interfacial tensions $\sigma_{p}^{ls}$ and $\sigma_{p}^{ls}$ of the prismal and 
basal facets, respectively, of such ice  crystals in supercooled  liquid water (ref.4, p.160) in Wulff's
relations allow one to obtain their approximate  aspect ratio $\kappa\approx 0.8$. 

Ice nucleation in supercooled water most likely  proceeds via the formation of nuclei of
metastable non-hexagonal ice modifications (reportedly cubic$^{40}$ or
stacking-disordered$^{15,41,42}$). However, one can  expect that the aspect ratio of such ice nuclei
will be also roughly equal to $0.8$, because it is determined via the ratio
$\sigma_b^{ls}/\sigma_p^{ls}$ in Wulff's relations which differs little between 
these three ice modifications due to relatively small differences between their corresponding interfacial
tensions.$^{4,15}$ We have thus assumed $\kappa\approx 0.8$.

In Figure 1 the combined ice nucleation rate $J=J(T,R)$ in  supercooled water droplets is plotted as a ratio
$J/J_v$ vs droplet radius $R$  at various $T$.  The nucleation rates $J$ were calculated according eq.(9), 
with the functions $h_{\lambda}(T)$ and $W_*(T)$ provided  by eqs.(7) and (5), respectively. For the
functions of temperature $\sigma^{ls}$, $\rho_i(T)$, and $\De q(T)$ in eqs.(5)-(7), we have adopted the 
Koop-Murray approximations.${35}$ 
The latter were also used in numerically solving eq.(8) (with respect to $u$) for the experimental data of 
Earle et al.$^{38}$ to construct the function $u(T)$, wherefor 
we obtained the following quadratic approximation
\beq u=u(T)=-1809.88+15.2553\,T-0.0321355\,T^2\;\;\;\;\;(233\mbox{ K}<T<240\mbox{ K}).\eeq

In Fig.1, the solid curve is for $T=235$ K, dashed curve for $T=237$ K, dash-dotted curve for  $T=239$ K, and
dotted curve for $T=239.3$ K. As clear, at any fixed $T$ the ratio $J/J_v$ is a monotonically decreasing
function of $R$. Starting from large values for smallest $R$, it sharply decreases with increasing $R$ and
asymptotically approaches $1$ as $R\rightarrow \infty$, when $J\rightarrow J_v$. At some droplet  radius
$R_{\mbox{\tiny s}}^{\mbox{\tiny v}}=R_{\mbox{\tiny s}}^{\mbox{\tiny v}}(T)$, the curve crosses the horizontal
line $J/J_v=2$.  In droplets of radius  $R=R_{\mbox{\tiny s}}^{\mbox{\tiny v}}$, the contributions from the
surface-stimulated mode and volume-based modes ($J_v^s$ and $J_v$, respectively) to the total rate of ice
nucleation are equal to each other,  $J_v^s=J_v$, so that $J=J_v^s+J_v=2J_v$.   Thus, the radius
$R=R_{\mbox{\tiny s}}^{\mbox{\tiny v}}$ determines the crossover radius for the transition of homogeneous ice
nucleation in supercooled droplets from being dominated by  the volume-based mode (with $J_v^s<J_v$ in
droplets of radii $R>R_{\mbox{\tiny s}}^{\mbox{\tiny v}}$) to being dominated by  the surface-stimulated mode
(with $J_v^s>J_v$ in droplets of radii $R<R_{\mbox{\tiny s}}^{\mbox{\tiny v}}$).

Figure 2 presents the $T$-dependence of the crossover radius $R^{\mbox{\tiny v}}_{\mbox{\tiny s}}$. 
In the range $233$ K $<T<239.5$ K  
the function $R^{\mbox{\tiny v}}_{\mbox{\tiny s}}=R^{\mbox{\tiny v}}_{\mbox{\tiny s}}(T)$ is 
non-monotonic and non-trivial. One can observe a sharp increase in 
$R^{\mbox{\tiny v}}_{\mbox{\tiny s}}$ as $T$ approaches $239.5$ K from left, as well as a local maximum of
about $70$ $\mu$m at $T\approx 233.5$ K. However (as seen in the inset in Fig.2), 
in the most interesting temperature range (for freezing of atmospheric droplets) $234.5$
K$<T<238.5$ K the crossover radius $R^{\mbox{\tiny v}}_{\mbox{\tiny s}}$ is in the range from $2$ $\mu$m to
$7$ $\mu$m, attaining its global minimum of about $2\;\mu$m for temperatures from $236$ K to $237.5$ K. 

The dependence of the combined homogeneous ice nucleation rate $J$ in supercooled water droplets (calculated
using eq.(9))  on the droplet radius $R$ and temperature $T$ is illustrated in Figure 3. The rate  $J$ as a
function of $R$ at constant $T$ is plotted in Fig.3a (dark solid curve is for $T=235$ K, dotted for $T=236$ K,
dashed for $T=237$ K, light solid for $T=238$ K, and dash-dotted for $T=239$ K). At any
temperature, $J(T,R)\rightarrow J_v(T)$ as $R\rightarrow \infty$, i.e.,  in large droplets  the
contribution from the surface-stimulated mode to $J$ tends to zero. For such (large) droplets, the 
$T$-dependence of $J(T,R)$ reduces to the $T$-dependence of its contribution from the  volume-based mode,
$J_v(T)$, given by eq.(4). 

These results are further illustrated   in Figure 3b, showing $J$ as a function of $T$ for droplets of a given
radius $R$;   different curves correspond to different $R$'s as indicated in the Figure panel. The lowest
curve represents infinitely large droplets, i.e., the function $J(T,\infty)=J_v(T)$. The smaller the
droplet, the larger the contribution $J_v^s$ from the surface-stimulated mode to  $J$, and it can exceed
the contribution $J_v$ from the volume-based mode by two orders of magnitude for 
$R\lesssim 0.5\;\mu$m.

Figure 4 shows the temperature dependence of the  ice-air interfacial tension $\sigma_{\lambda}^{sv}$ of the
basal  facet $\lambda$ of ice nuclei  obtained by using the experimental data$^{38}$  
for 
$J^{\mbox{\tiny exp}}_{\mbox{\tiny small}}$ for small  droplets of radii $R=1,\;1.7$, and $2.9$ $\mu$m. 
The results for $\sigma_{\lambda}^{sv}$ at each
temperature represent the average over three values extracted from each experimental set  
(for droplets of three different radii). Despite relatively large uncertainties in 
reported$^{38}$ temperatures $T$ and nucleation rates $J^{\mbox{\tiny exp}}_{\mbox{\tiny small}}$, 
one can observe relatively smooth, monotonic, linear-like dependence of 
$\sigma_{\lambda}^{sv}$ on $T$, with the interfacial tension 
$\sigma_{\lambda}^{sv}$ increasing from about $89$ dyn/cm to $91$ dyn/cm
as the temperature increases from $T=234.8$ to $T=236.2$.

\subsection{Concluding remarks}
Thus, we have proposed  a relatively simple approximation  for the total rate of homogeneous ice nucleation
rate in supercooled water  droplets as a function of both droplet radius and temperature. It contains the
contributions from both the surface-stimulated and volume-based modes of ice nucleation and thus more
adequately describes the freezing of  supercooled water droplets in cirrus clouds. This approximation can 
improve the accuracy of both regional and global climate models once implemented therein (in a parameterized
form). 

The crucial idea  of our method is that, even in the  surface-stimulated mode, when a crystal nucleus
forms with one of its facets at  the droplet surface, it initially emerges (as a
subcritical cluster) {\em homogeneously} in the  subsurface layer, not ``pseudo-heterogeneously"  at the
surface.   This mode is negligible for large droplets, but becomes increasingly  important with decreasing
droplet size and is dominant in small droplets. We have proposed a definition of a crossover radius for the
transition of homogeneous ice nucleation in droplets from the volume-based mode to the
surface-stimulated one and determined its temperature dependence.  We have also determined the solid-vapor
interfacial tension of the basal facets of ice crystals (assumed to be shaped as hexagonal prisms) 
using experimental data on ice nucleation rates in small droplets.



\section*{References}
\begin{list}{}{\labelwidth 0cm \itemindent-\leftmargin} 

\item $(1)$F.  Franks, {\it Water: A matrix for life}; Royal Society of Chemistry: Cambridge, 2000.
\item $(2)$ The intergovernmental panel on climate change, Fourth assessment report: Climate change 2007 (AR4)
working group I report: ``The physical basis", 2007.
\item $(3$ N. H. Fletcher, {\it The physics of rainclouds}, University Press, Cambridge, 1962.
\item $(4)$ H. R. Pruppacher and J. D. Klett., {\it Microphysics of clouds and precipitation}, 
Kluwer Academic Publishers, Norwell, MA, 1997. 
\item $(5)$ R. B. Lawson, B. Baker, B. Pilson, and Q. Mo, 
{\it J. Atmos. Sci.}, 2006, {\bf 63}, 3186-3203.
\item $(6)$ K. Liou, 
{\it Mon. Weather. Rev.} 1986, {\it 114}, 1167-1199.
\item $(7)$ S. K. Cox, {\it J. Atmos. Sci.}, 1971, {\bf 28}, 1513-1515.
\item $(8)$ A. Ravishankara, {\it Science}, 1997, {\bf 276}, 1058-1065.
\item $(9)$ E. J. Jensen, et al. 
{\it Geophys. Res. Lett.}, 1998, {\bf 25}, 1363-1366.
\item $(10)$ D. C. Rogers, P. J. Demott, S. Kreidenweis, and  Y. Chen, 
{\it Geophys. Res. Lett.}, 1998, {\bf 25}, 1383-1386. 
\item $(11)$ A. J. Heymsfield and L. M. Miloshevich, 
{\it J. Atmos. Sci.}, 1993, {\bf 50}, 2335.
\item $(12)$ A. Tabazadeh, E. J. Jensen, O. B. Toon, K. Drdla, and M. R. Schoeberl, {\it Science}, 2001,
{\bf 291}, 2591. 
\item $(13)$ D. Salcedo, L. T. Molina, and M. J. Molina,  {\it J. Phys. Chem.}, 2001, {\bf 105}, 1433.
\item $(14)$ W. C. Thuman and E. Robinson, 
{\it J. Meteorol.}, 1954, {\bf 11}, 151-156.
\item $(15)$ T. Koop and B. J. Murray,  {\it J. Chem. Phys.}, 2016, {\bf 145}, 211915.
\item $(16)$ M. V\"olmer,  {\it Kinetik der phasenbildung}, Teodor Steinkopff, Dresden und Leipzig, 1939.
\item $(17)$ D. Turnbull and J. C. Fisher, 
{\it J. Chem. Phys.}, 1949, {\bf 17}, 71-73.
\item $(18)$ A. Tabazadeh, Y. S. Djikaev, P. Hamill, and H. Reiss, 
{\it J. Phys. Chem. A}, 2002, {\bf 106}, 10238-10246.
\item $(19)$ A. Tabazadeh, Y. S. Djikaev, and H. Reiss, 
{\it Proc. Natl. Acad. Sci. USA}, 2002, {\bf 99}, 15873.
\item $(20)$ Y. S. Djikaev,  A. Tabazadeh, P. Hamill, and H. Reiss, 
{\it J.Phys.Chem. A}, 2002, {\bf 106}, 10247.
\item $(21)$ Y. S. Djikaev, A. Tabazadeh, and H. Reiss,  
{\it J.Chem.Phys.}, 2003, {\bf 118}, 6572-6581.
\item $(22)$ R. Defay, I. Prigogine, A. Bellemans, and D. H. Everett, 
{\it Surface tension and adsorption}, John Wiley, New York, 1966.
\item $(23)$E. Ruckenstein and G. Berim, 
{\it Kinetic theory of nucleation}, (CRC, New York, 2016).
\item $(24)$ B. Mutaftschiev and J. Zell, 
{\it Surf. Sci.}, 1968, {\bf 12}, 317.
\item $(25)$ G. Grange, R. Landers, and B. Mutaftshiev, 
{\it Surf. Sci.}, 1976, {\bf 54}, 445-462.
\item $(26)$ D. Chatain and P. Wynblatt, in {\em Dynamics of crystal surfaces and interfaces},   
Ed. P. M. Duxbury and T.J.Pence, 53-58; Springer, NY, 2002.
\item $(27)$ M. Elbaum, S. G. Lipson, and  J. G. Dash, 
{\it J. Cryst. Growth}, 1993, {\bf 129}, 491-505.
\item $(28)$ Y. S. Djikaev, 
{\it J. Phys. Chem. A}, 2008, {\bf 112}, 6592-6600.
\item $(29)$ Y. Djikaev and E. Ruckenstein, 
{\it Physica A}, 2008, {\bf 387}, 134-144. 
\item $(30)$ X. Wei, P. B. Miranda, and Y.R.Shen, 
{\it Phys. Rev. Lett.}, 2001, {\bf 86}, 1554-1557. 
\item $(31)$ X. Wei, Y. R. Shen, 
{\it Applied Phys. B}, 2002, {\bf 74}, 617-620.
\item $(32)$ T. Kuhn, M. E. Earle, A. F. Khalizov, and J. J. Sloan, 
{\it Atmos. Chem. Phys.}, 2011, {\bf 11}, 2853-2861.
\item $(33)$ S. Toxvaerd, N. Larsen, and J. C. Dyre, 
{\it J. Phys. Chem. C}, 2011, {\bf  115},  12808-12814.
\item $(34)$ Y. S. Djikaev and  E. Ruckenstein, 
{\it J. Phys. Chem. C}, 2016, {\bf 120}, 28031-28037.
\item $(35)$ Y. S. Djikaev and  E. Ruckenstein, 
{\it J. Chem. Phys.}, 2017, {\bf 058712JCP}, (in print). 
\item $(36)$ A. R. MacKenzie, M. Kulmala, A. Laaksonen, T. Vesala, 
{\it J. Geophys. Res.}, 1995, {\bf 100}, 11275-11288. 
\item $(37)$ J. Hruby, V. Vins, R. Mares, J. Hykl, and J. Kalova, 
{\it J. Phys. Chem. Lett.}, 2014, {\bf 5}, 425-428.
\item $(38)$ M. E. Earle, T. Kuhn, A. F. Khalizov, and J. J. Sloan,  
{\it Atmos. Chem. Phys.}, 2010, {\bf 10}, 7945-7961.
\item $(39)$ T. Kobayashi, {\it Phil. Mag.}, 1961, {\bf 6}, 1363.
\item $(40)$ J. Huang and L. Bartell, 
{\it J. Phys. Chem.} 1995, {\bf 99}, 3924-3931.
\item $(41)$ T. Li, D. Donadio, G. Russo, G. Galli, 
{\it Phys. Chem. Chem. Phys.}, 2011, {\bf 13}, 19807.
\item $(42)$ T. L. Malkin, B. J. Murray, A. V. Brukhno, J. Anwar, C. G. Salzmann, 
{\it Proc. Natl. Acad. Sci. USA}, 2012, {\bf 109}, 1041.

\end{list}
\newpage 
\subsection*{Captions} 
to  Figures 1 to 4 of the manuscript {\sc
``Dependence of homogeneous crystal nucleation in water droplets on their radii and its implication for 
modeling the formation of ice particles in cirrus clouds."
}  by {\bf  Y. S. Djikaev} and {\bf E. Ruckenstein}. 
\subsubsection*{}
\vspace{-1.0cm}   
Figure 1. The dependence of the ratio $J(T,R)/J_v(T)$ in supercooled water droplets  
on the droplet radius $R$  
at three different temperatures: $235$ K (solid curve), $237$ K (dashed curve), $239$ K (dash-dotted
curve), and $239.3$ K  (dotted curve). 
\vspace{0.3cm}\\   
Figure 2. The temperature dependence of the crossover radius $R^{\mbox{\tiny v}}_{\mbox{\tiny s}}$
for the transition of ice nucleation in supercooled water 
droplets from the volume-based mode (at $R>R^{\mbox{\tiny
v}}_{\mbox{\tiny s}}$) to the surface-stimulated mode (at $R<R^{\mbox{\tiny
v}}_{\mbox{\tiny s}}$). 
\vspace{0.3cm}\\
Figure 3. The dependence of the combined ice nucleation rate $J$ in supercooled water droplets  
on the droplet radius $R$ and temperature $T$. a) $J$ as a function of $R$ at constant $T$;  
different curves correspond to different  temperatures as marked. 
b) $J$ as a function of $T$ for droplets of given radius $R$;  
different curves correspond to different droplet radii $R$ as indicated in the Figure panel. 
\vspace{0.3cm}\\ 
Figure 4. The temperature dependence of the 
ice-air interfacial tension $\sigma_{\lambda}^{sv}$ of the basal 
facet $\lambda$ of ice crystal
nuclei  obtained by using the experimental data$^{38}$ for the homogeneous ice nucleation rates 
$J^{\mbox{\tiny exp}}_{\mbox{\tiny small}}$ in small supercooled water 
droplets of radii $1$ $\mu$m, $1.7$ $\mu$m, and $2.9$ $\mu$m. 

\newpage
\begin{figure}[htp]
\begin{center}\vspace{2cm}
\includegraphics[width=8.3cm]{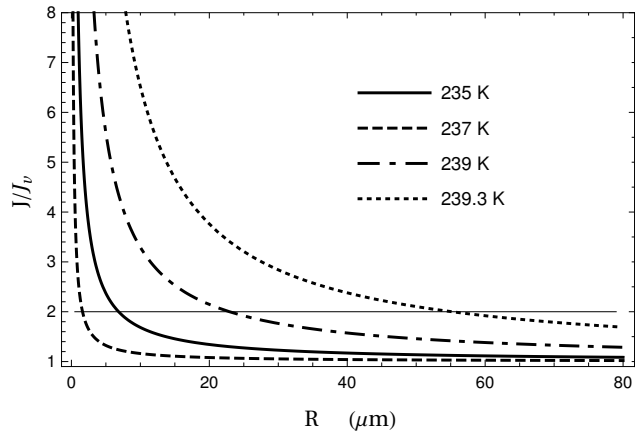}\\ [1.cm]
\renewcommand{\baselinestretch}{0.8}\normalsize
\caption{\small The dependence of the ratio $J(T,R)/J_v(T)$ in supercooled water droplets  
on the droplet radius $R$  
at three different temperatures: $235$ K (solid curve), $237$ K (dashed curve), $239$ K (dash-dotted
curve), and $239.3$ K  (dotted curve).}
\end{center}
\end{figure} 

\newpage
\begin{figure}[htp]
\begin{center}\vspace{2cm}
\includegraphics[width=8.3cm]{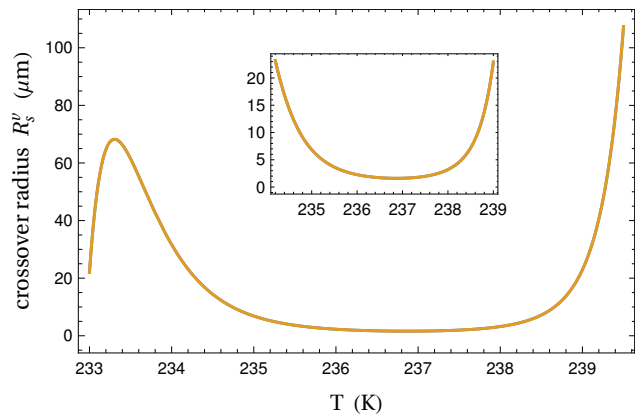}\\ [1.cm]
\renewcommand{\baselinestretch}{0.8}\normalsize
\caption{\small The temperature dependence of the crossover radius $R^{\mbox{\tiny v}}_{\mbox{\tiny s}}$
for the transition of ice nucleation in supercooled water 
droplets from the volume-based mode (at $R>R^{\mbox{\tiny
v}}_{\mbox{\tiny s}}$) to the surface-stimulated mode (at $R<R^{\mbox{\tiny
v}}_{\mbox{\tiny s}}$).}
\end{center}
\end{figure} 

\newpage
\begin{figure}[htp]\vspace{1cm}
	      \begin{center}
$$
\begin{array}{c@{\hspace{0.3cm}}c} 
              \leavevmode
      	      \vspace{0.3cm}
	\leavevmode\hbox{a) \vspace{0.3cm}} &   
\includegraphics[width=8.3cm]{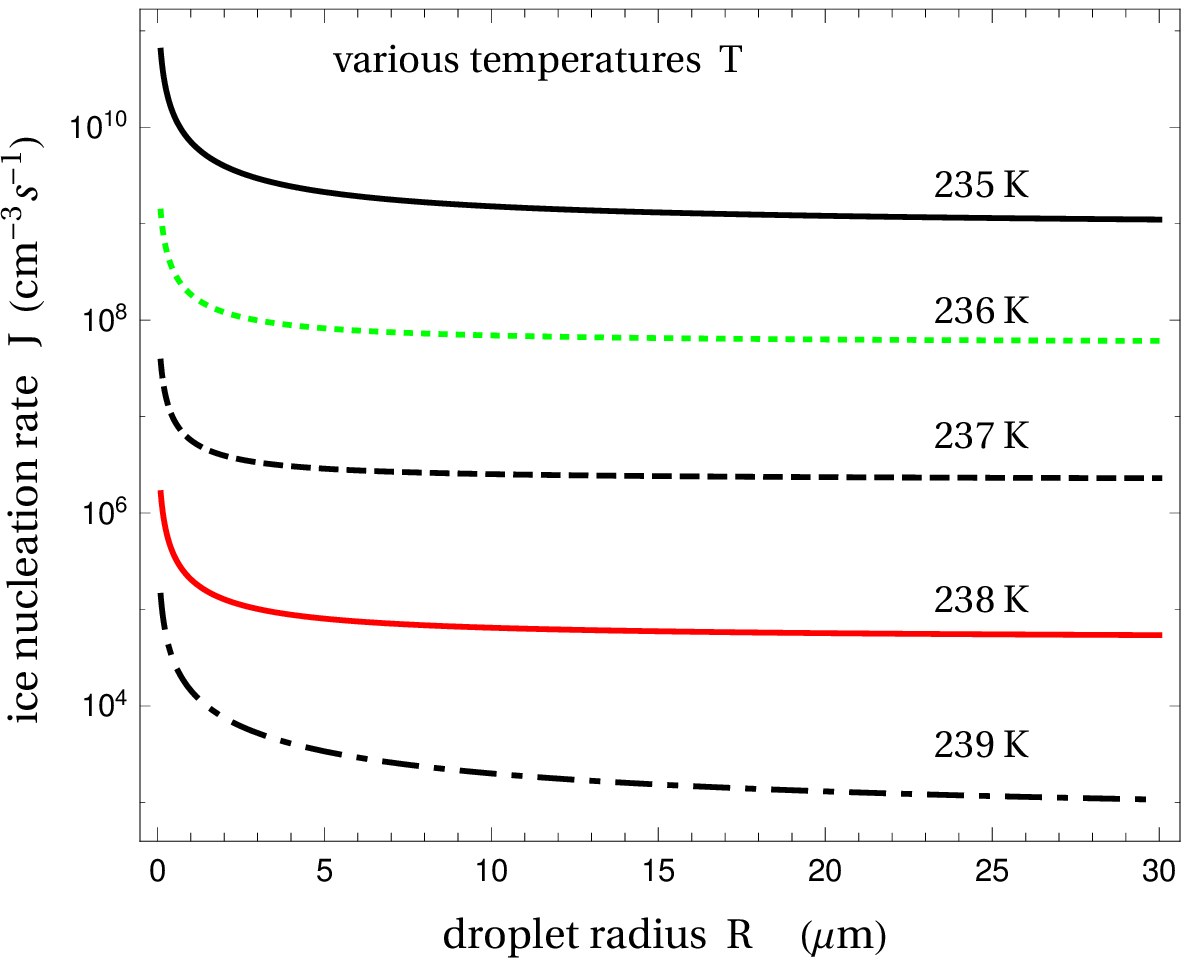}\\ [0.3cm] 
      	      \vspace{1.0cm}
	\leavevmode\hbox{b) \vspace{0.3cm}} &  
      	      \vspace{0.0cm}
\includegraphics[width=8.3cm]{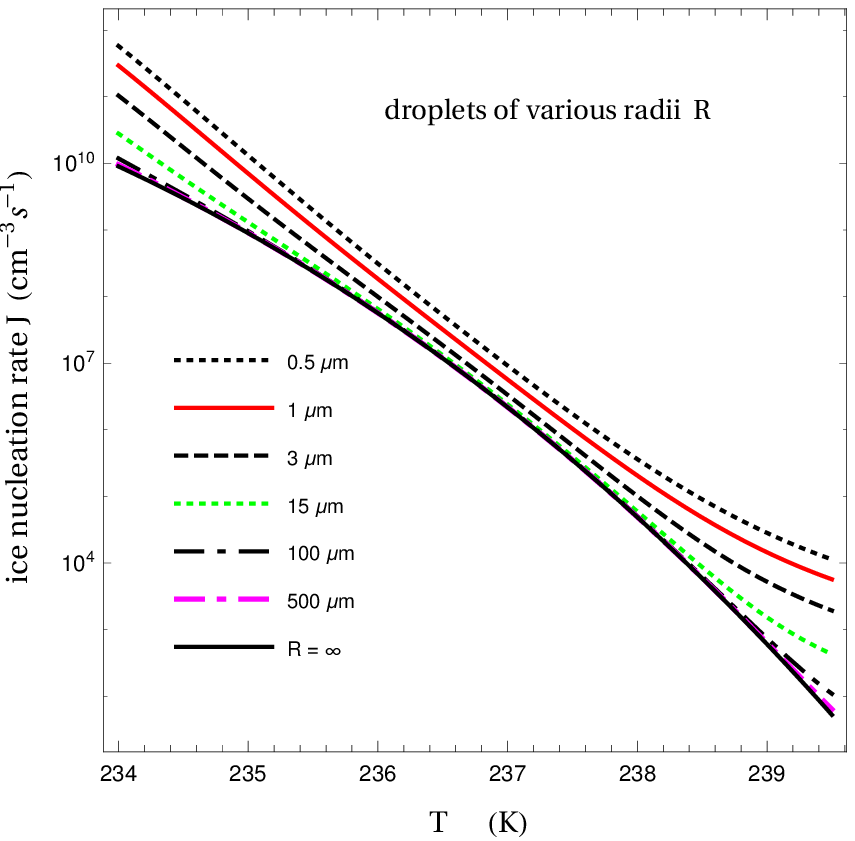}\\ [0.3cm] 
      	      \vspace{-0.8cm}
\end{array}  
$$  

	      \end{center} 
\renewcommand{\baselinestretch}{0.8}\normalsize
            \caption{\small The dependence of the combined ice nucleation rate $J$ in supercooled water droplets  
on the droplet radius $R$ and temperature $T$. a) $J$ as a function of $R$ at constant $T$;  
different curves correspond to different  temperatures as marked. 
b) $J$ as a function of $T$ for droplets of given radius $R$;  
different curves correspond to different droplet radii $R$ as indicated in the Figure panel. } 
\end{figure}

\newpage
\begin{figure}[htp]
\begin{center}\vspace{-1cm}
\includegraphics[width=8.3cm]{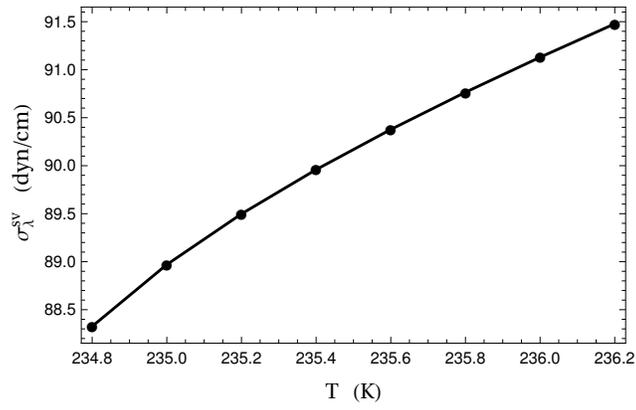}\\ [1.cm]
\renewcommand{\baselinestretch}{0.8}\normalsize
\caption{\small The temperature dependence of the 
ice-air interfacial tension $\sigma_{\lambda}^{sv}$ of the basal 
facet $\lambda$ of ice crystal
nuclei  obtained by using the experimental data$^{38}$ for the homogeneous ice nucleation rates 
$J^{\mbox{\tiny exp}}_{\mbox{\tiny small}}$ in small supercooled water 
droplets of radii $1$ $\mu$m, $1.7$ $\mu$m, and $2.9$ $\mu$m.}
\end{center}
\end{figure}

\end{document}